\documentclass[useAMS,usenatbib]{mn2e}
\usepackage{amssymb}
\usepackage{graphics}

\def\eqref#1{equation~(\ref{#1})}
\def\eqrefs#1#2{equations~(\ref{#1})-(\ref{#2})}
\def\apref#1{Appendix~\ref{#1}}
\def\scref#1{Section~\ref{#1}}
\def\binom#1#2{{#1 \choose #2}}

\title[Linearized Equations of the $N$-body Problem]{Solving Linearized Equations of the $N$-body Problem Using the Lie-integration Method}
\author[A. P\'al and \'A. S\"uli]{Andr\'as P\'al%
$^{1}$\thanks{E-mail:
apal@szofi.elte.hu (AP); a.suli@astro.elte.hu (\'AS)} and 
\'Aron S\"uli$^{1}$\\
$^{1}$Department of Astronomy, Lor\'and E\"otv\"os University, P\'azm\'any P\'eter s\'et\'any 1/A, Budapest H-1117, Hungary}

\begin{document}

\date{Accepted 2007 July 18. Received 2007 July 13; in original form 2007 May 25}

\pagerange{\pageref{firstpage}--\pageref{lastpage}} \pubyear{2007}

\maketitle

\label{firstpage}

\begin{abstract}
Several integration schemes exits to solve the equations of motion of the
$N$-body problem. The Lie-integration method is based on the idea to solve
ordinary differential equations with Lie-series. In the 1980s this method was
applied for the $N$-body problem by giving the recurrence formula for the
calculation of the Lie-terms. The aim of this works is to present the
recurrence formulae for the linearized equations of motion of $N$-body 
systems. We prove a lemma which greatly simplifies the derivation of 
the recurrence formulae for the linearized equations if the recurrence 
formulae for the equations of motions are known. The Lie-integrator is 
compared with other well-known methods. The optimal
step size and order of the Lie-integrator are calculated. It is shown that a
fine-tuned Lie-integrator can be 30\%-40\% faster than other 
integration methods.
\end{abstract}


%%%%%%%%%%%%%%%%%%%%%%%%%%%%%%%%%%%%%%%%%%%%%%%%%%%%%%%%%%%%%%%%%%%%%%%%%%%%%%

\section{Introduction}

The classical problems of celestial mechanics are described by 
a system of ordinary differential equations (ODEs). The investigation
of the motions in the Solar System, exoplanetary systems, 
satellites around the Earth or other celestial objects are based on the
solutions of such ODEs. However, several modern analysis,
including many chaos detection methods require to solve 
the linearized equations of the problem.

The integration method based on the Lie-series \citep{grobner1967}
is widely used in celestial mechanics to solve ODEs
(see \citet{hanslmeier1984}, hereafter H\&D and articles 
referring to it). The basis of this method is to generate the coefficients 
of the Taylor expansion of the solution by using recurrence relations.
The principal application, i.e. the integration of the $N$-body problem 
is described in details in H\&D.

\subsection{Lie-integration}

Here we summarize the key points of this method of numerical
integration, using almost identical notations as used 
by \citet{hanslmeier1984}.

Let us write the differential equation to be solved as
\begin{equation}
\dot x_i=f_i(\mathbf{x}), \label{diffeq}
\end{equation}
where $\mathbf{x}\equiv (x_1,\dots,x_N)$ 
is an $\mathbb{R} \to \mathbb{R}^N$ and 
$\mathbf{f}\equiv (f_1,\dots,f_N)$ is an $\mathbb{R}^N \to \mathbb{R}^N$
continuous function and $N$ is the dimension of the vector $\mathbf{x}$
and the vector space where $\mathbf{f}$ maps from and maps to. 
Let us introduce the differential operator
\begin{equation}
D_i := \frac{\partial}{\partial x_i}\label{shortdiff},
\end{equation}
and the derivation
\begin{equation}
L_0 := \sum\limits_{i=1}^{N} f_i\frac{\partial}{\partial x_i}, \label{lieopdef}
\end{equation}
which is known as the Lie-derivation or Lie-operator. $L_0$ is a 
linear differential operator and one can apply Leibniz's rule,
\begin{equation}
L_0(ab)=aL_0(b) + bL_0(a)\label{leibnitz},
\end{equation}
where $a$ and $b$ are $\mathbb{R}^N \to \mathbb{R}^N$ differentiable
functions. It can easily be proven that the 
solution of \eqref{diffeq} at a given instance $t+\Delta t$
is formally
\begin{equation}
\mathbf{x}(t+\Delta t) = \exp\left(\Delta t\cdot L_0\right)\mathbf{x}(t), \label{liesol1}
\end{equation}
where 
\begin{equation}
\exp\left(\Delta t\cdot L_0\right)=
\sum\limits_{k=0}^\infty \frac{\Delta t^k}{k!}L_0^k=
\sum\limits_{k=0}^\infty \frac{\Delta t^k}{k!}
\left(\sum\limits_{i=1}^N f_iD_i\right)^k. \label{liesol2}
\end{equation}
The method of Lie-integration is finite approximation 
of the sum in the right-hand side of 
\eqref{liesol2}, up to the order of $M$, namely
\begin{equation}
\mathbf{x}(t+\Delta t)\approx\left(\sum\limits_{k=0}^{M} \frac{\Delta t^k}{k!}L_0^k\right)
\mathbf{x}(t)=\sum\limits_{k=0}^{M} \frac{\Delta t^k}{k!}\left(L_0^k\mathbf{x}(t)\right).\label{lienumint}
\end{equation}
The proof of \eqref{liesol1} and other related properties of 
the Lie-derivation can be found in \citet{grobner1967} or 
\citet{hanslmeier1984}.

In spite of the fact that the Lie-derivatives can analytically be calculated
up to arbitrary order, the formulae yielded by these expansions are highly
complicated even if all kinds of new variables are introduced
(see e.g. equations (19d) or (19e) in H\&D at page 204). A definitely more
efficient way to evaluate the Lie-derivatives is to find a set of
recurrence relations. These relations allow us to express the $(n+1)$th
Lie-derivative, e.g. $L^{n+1}\mathbf{x}$ as the function of the derivatives
with lower order, namely $L^{j}\mathbf{x}$ where $0\le j\le n$. The initialization
of such a recurrence relation is evident, because 
$L^0\mathbf{x}\equiv\mathbf{x}$. We note that
in several applications, well-chosen auxiliary variables have to be introduced 
to gain a compact set of recurrence relations which can efficiently be
evaluated.

\subsection{The importance of linearized equations}

Wide range of problems related to celestial mechanics require to 
solve simultaneously the linearized form of the original equations too.
The numerous experiments conducted in the last decades show that chaotic
behaviour is typical and already occurs in simple but nonlinear systems. This
finding throws completely new light upon these systems and the study of chaotic
behaviour became of high concern. A major part of the frontline research
focuses on the structure of the phase space, therefore the problem to 
separate ordered and chaotic motion in systems, which posses only a few 
degrees of freedom and are described by ODEs,
has become a fundamental task in a wide area of modern research. The phase
space of these nonlinear systems can not be described by the known 
mathematical tools. To map the phase space and study the chaotic 
behaviour of a given system fast and reliable numerical tools are needed. 
These tools are extremely useful in those cases when the inspected 
dynamical system has more than two degrees of freedom and
accordingly its phase space cannot be explored in a direct way or 
the classical method of surface of section (SoS) can not be applied which is 
widely used in the case of conservative systems with two degrees and 
freedom. The basic idea of the method of SoS was 
invented by \citet{poincare1899} and its application was renewed by
\citet{henon1964}.

The mathematical foundation of the theory of Lyapunov Characteristic Exponents
(LCEs) is approximately of the same age as the SoS and arose progressively in
the literature. The use of such exponents dates back to \citet{lyapunov1907},
but was first applied by \citet{oseledec1968} to characterize trajectories.
\citet{henon1964} found that in an integrable region of the phase space of a
dynamical system nearby orbits diverge linearly whereas in a chaotic region
they diverge exponentially. The LCEs express these facts in a precise form
and many papers were devoted to the application of LCEs in several nonlinear
problems.

Unfortunately both methods have a serious drawback. To compute the LCEs the
equations have to integrate for infinity, which is numerically impossible. 
The method of SoS becomes hard to handle and greatly deceiving for systems 
with more than two degrees of freedom. To overcome these problems was the 
main motivation in the 1980s that initiated the research to develop new 
numerical methods to characterize the stochasticity of the trajectories
in the phase space in short time-span and in arbitrary dimension.
The developed methods can be classified in two groups:
one group consists of the methods which are based on the analysis of the
orbits, (e.g.~SoS or frequency analysis, see \citet{laskar1990}),
the other one is based on the time evolution of the tangent vector, i.e.~the
solution of the linearized equations of motion (e.g.~LCE). There
are complete software packages designed to analyse systems of celestial
mechanics, both for for general integration of motion 
\citep[e.g. Mercury6, see][]{chambers1999} and for solving linearized
equations and calculating LCEs \citep[ORBIT9, see][]{milani1988}.
We also have to mention that there are several improved chaos detection methods which
are based on the solution of the linearized equations. Instead of a 
complete review, we only mention two of them: the method
of Fast Lyapunov Indicators \citep[FLIs, see][]{froeschle1997} and
the method of Mean Exponential Growth of Nearby Orbits 
\citep[MEGNO, see][]{cincotta2000,godziewski2001}.

The aim of this paper is 
to present a lemma which advances the derivation of the same kind of
recurrence relations for the linearized equations. We present these
relations for certain classical dynamic systems: for
the general $N$-body problem and for the $N$-body problem in the
reference frame of one of the bodies. In the last section we compare
the efficiency of this method with well-known other ones.

\section{Linearized equations}

The chaos indicators mentioned in the previous section can be obtained if
the linearized form of the equations of motion is solved. 
The solution of the linearized equations is an $\mathbf{\xi}\equiv
(\xi_1,\dots,\xi_N): \mathbb{R}\to\mathbb{R}^N$ function having the same
dimension as the equations of motion has. The linearized
equations of \eqref{diffeq} (any ODE can be written in this form)
can be written as
\begin{equation}
\dot \xi_i=\sum\limits_{m=1}^{N}\xi_m \frac{\partial f_i(\mathbf{x})}{\partial x_m}, \label{lineq}
\end{equation}
where the variables $\xi_i\equiv\xi_i(t):\mathbb{R}\to\mathbb{R}$ are 
the so-called linearized variables
and $\mathbf{x}\equiv\mathbf{x}(t)$ is the solution of \eqref{diffeq}. 
Equation (\ref{lineq}) is linear in $\mathbf{\xi}$, therefore if $\xi^{(1)}_i(t)$
and $\xi^{(2)}_i(t)$ are two independent solutions, 
then $\alpha\xi^{(1)}_i(t)+\beta\xi^{(2)}_i(t)$
is also a solution. Using the Einstein summation convention \eqref{lineq}
can be written in a more compact form:
\begin{equation}
\dot \xi_i=\xi_m D_m f_i. \label{shortlindiff}
\end{equation}

\subsection{Lie-derivatives of the linearized equations}

Introducing the differential operator
\begin{equation}
\partial_i := \frac{\partial}{\partial\xi_i}\label{lindiffoper},
\end{equation}
the coupled system of equations (both the original and the linearized) is
\begin{eqnarray}
\dot x_i & = & f_i, \label{fulleq1} \\
\dot \xi_i & =  & \xi_m D_m f_i,  \label{fulleq2}
\end{eqnarray}
and the Lie-operator of \eqrefs{fulleq1}{fulleq2} is 
\begin{equation}
L = L_0 + L_\ell = f_i D_i + \xi_m D_m f_i \partial_i. \label{liefulleq}
\end{equation}

\paragraph*{Lemma}
Using the same notations as above the Lie-deriva\-tives of $\xi_k$
can be written as
\begin{equation}
L^n\xi_k=\xi_mD_mL^nx_k=\xi_mD_mL_0^nx_k\label{linliederiv}.
\end{equation}

\paragraph*{Proof}
Obviously, \eqref{linliederiv} is true for $n=0$:
\begin{equation}
D_mL^0x_k=D_mx_k=\delta_{mk}, 
\end{equation}
hence
\begin{equation}
\xi_mD_mL^0x_k=\xi_m\delta_{mk}=\xi_k.
\end{equation}
Let us suppose that it is true for all $0\le j\le n$ and calculate the
$(n+1)$th Lie-derivative of $\xi_k$:
\begin{eqnarray}
L^{n+1}\xi_k & = & L\left(\xi_mD_mL^nx_k\right) = \nonumber \\
& = & \left(f_iD_i+\xi_jD_jf_i\partial_i\right)\left(\xi_mD_mL^nx_k\right) = \nonumber\\
& = & f_iD_i\xi_mD_mL^nx_k + \nonumber \\
& &  + \xi_j(D_jf_i)[\delta_{im}D_mL^nx_k + \nonumber \\
& &  + \xi_mD_m\partial_iL^nx_k]. \label{linrec1}
\end{eqnarray}
Here the term $\xi_mD_m\partial_iL^nx_k$ equals to zero, because $x_k$ and
$L^nx_k$ for all $n\ge 0$ do not depend on $\xi$. Therefore,
\begin{eqnarray}
L^{n+1}\xi_k & = & f_iD_i\xi_mD_mL^nx_k+\xi_j(D_jf_i)D_iL^nx_k = \nonumber \\
& = & \xi_mf_iD_mD_iL^nx_k + \xi_m(D_mf_i)(D_iL^nx_k) = \nonumber \\
& = & \xi_m \left( f_iD_m + D_mf_i \right)(D_iL^nx_k) = \nonumber \\
& = & \xi_m D_m (f_iD_i)(L^nx_k) = \nonumber \\
& = & \xi_m D_m L ( L^nx_k ) = \xi_m D_m L^{n+1}x_k \nonumber \\
& = & \xi_m D_m L_0^{n+1}x_k. \label{linrec2}
\end{eqnarray}
We have applied Young's theorem, namely
\begin{equation}
D_mD_i = D_iD_m,
\end{equation}
and Leibniz rule, 
\begin{eqnarray}
D_m(f_iD_i)X & = & D_mf_i(D_iX) = \nonumber \\
 & = & f_i(D_mD_iX)+(D_mf_i)(D_iX),
\end{eqnarray}
where $X$ can be an arbitrary function of $\mathbf{x}$, in \eqref{linrec2}
$X\equiv L^n x_k$. Therefore \eqref{linrec2} is the same relation
for $n+1$, as \eqref{linliederiv} for $n$.
Continuing the scheme described above, \eqref{linliederiv}
can be proven for all positive integer values of $n$. \hfill $\blacksquare$

\subsection{An example: applying to the H\'enon-Heiles system}

Demonstrating the power of the lemma proven in the previous
subsection we derive the recurrence relations for the equation of the 
H\'enon-Heiles dynamical system and its linearized form. The H\'enon-Heiles
system is one of the simplest Hamiltonian systems which shows chaotic
behaviour under certain initial conditions \citep[see][]{henon1964}.

The equations of motion are derived from the Hamiltonian function
\begin{eqnarray}
H(x,y;v,w) & = & \frac{1}{2}\left(x^2+y^2+2x^2y-\frac23y^3\right)+ \nonumber\\
	& & +\frac12\left(v^2+w^2\right),
\end{eqnarray}
where $\dot x=v$ and $\dot y=w$. 
The equations of motion are
\begin{eqnarray}
\dot x & = & v, \label{eqhenon1} \\
\dot y & = & w, \label{eqhenon2} \\
\dot v & = & -x-2xy, \label{eqhenon3} \\
\dot w & = & -y-x^2+y^2, \label{eqhenon4}
\end{eqnarray}
and the Lie-operator of this system of equations is
\begin{equation}
L_0 = v\partial_x + w\partial_y + (-x-2xy)\partial_v + (-y-x^2+y^2)\partial_w,
\end{equation}
according to \eqref{lieopdef}.
It can easily be shown that the recurrence relations of the equations
(\ref{eqhenon1})~-~(\ref{eqhenon4}) are the following,
\begin{eqnarray}
L_0^{n+1}x & = &  L_0^nv, \\
L_0^{n+1}y & = &  L_0^nw, \\
L_0^{n+1}v & = & -L_0^nx-2\sum\limits_{k=0}^n\binom{n}{k}L_0^kxL_0^{n-k}y, \\
L_0^{n+1}w & = & -L_0^ny-\sum\limits_{k=0}^n\binom{n}{k}\left(L_0^kxL_0^{n-k}x-\right. \nonumber \\
	 & & \left.-L_0^kyL_0^{n-k}y\right).
\end{eqnarray}
Let us denote the linearized variables related to $x$, $y$, $v$ and $w$
by $\xi$, $\eta$, $\phi$ and $\rho$, respectively. According to 
\eqref{linliederiv} the Lie-derivatives of these variables are
\begin{eqnarray}
L^n\xi & = & \xi_mD_mL^nx,  \\
L^n\eta & = & \xi_mD_mL^ny,  \\
L^n\phi & = & \xi_mD_mL^nv, \\
L^n\rho & = & \xi_mD_mL^nw,
\end{eqnarray}
where $\xi_1\equiv\xi$, $\xi_2\equiv\eta$, $\xi_3\equiv\phi$ and 
$\xi_4\equiv\rho$. The pure recurrence relations can be almost automatically
derived. For the first two variables it is
evidently
\begin{eqnarray}
L^{n+1}\xi & = & \xi_mD_mL^{n+1}x = \xi_mD_mL^nv=L^n\phi,  \\
L^{n+1}\eta & = & \xi_mD_mL^{n+1}y = \xi_mD_mL^nw=L^n\rho. 
\end{eqnarray}
For the third variable one gets
\begin{eqnarray}
L^{n+1}\phi & = & \xi_mD_mL^{n+1}v = \nonumber \\
	& = & \xi_mD_m\left(-L^nx-2\sum\limits_{k=0}^n\binom{n}{k}L^kxL^{n-k}y\right)= \nonumber \\
	&=& - \xi_mD_mL^nx - \nonumber \\
	& & - 2\sum\limits_{k=0}^n\xi_m\binom{n}{k}D_m\left[L^kxL^{n-k}y\right] = \nonumber \\
	&=& - L^n\xi - 2\sum\limits_{k=0}^n\binom{n}{k}\xi_m\left[(D_mL^kx)(L^{n-k}y)+\right. \nonumber \\
	& & \left.+(L^kx)(D_mL^{n-k}y)\right] = \nonumber \\
	&=& - L^n\xi - 2\sum\limits_{k=0}^n\binom{n}{k}\left[(\xi_mD_mL^kx)(L^{n-k}y)+\right. \nonumber \\
	& & \left.+(L^kx)(\xi_mD_mL^{n-k}y)\right] = \nonumber \\
	&=& - L^n\xi - 2\sum\limits_{k=0}^n\binom{n}{k}\left[L^k\xi L^{n-k}y+\right. \nonumber \\
	& & \left.+L^kxL^{n-k}\eta\right].
\end{eqnarray}
The same procedure can be performed for $\rho$ and the result is
\begin{eqnarray}
L^{n+1}\rho & = &  - L^n\eta - \sum\limits_{k=0}^n\binom{n}{k}\left[L^k\xi L^{n-k}x+L^kx L^{n-k}\xi-\right.\nonumber\\
& & \left.-L^k\eta L^{n-k}y-L^kyL^{n-k}\eta\right]. 
\end{eqnarray}

%%%%%%%%%%%%%%%%%%%%%%%%%%%%%%%%%%%%%%%%%%%%%%%%%%%%%%%%%%%%%%%%%%%%%%%%%%%%%%

\section{The Lie-derivatives for the $N$-body problem and its linearized form}

Let us have $K$ point masses $m_i$ ($i=1,\dots,K$) moving
under the mutual gravitational attraction described by Newton's 
universal law of gravity. The coordinates
and the velocities of these particles are denoted by 
$x_{im}$ and $v_{im}$, where
$m$ is the index for the spatial dimension ($m=1,2,3$). In the 
following sections we denote the bodies by indices $i$, $j$, $k$, \dots
and the spatial indices by $m$, $n$, $p$, \dots, therefore the 
Einstein summation convention should be performed between the appropriate
limits, which is not explicitly noted everywhere.

Following H\&D, we present the derivation of the recurrence formulae for the
Lie-derivatives. The whole calculation is presented in 
\apref{appendixrecurrnbody}. We note that with different types of notations
the calculation can also be found in H\&D, some steps of the 
derivation should be emphasized for further calculations of the 
linearized equations.

\subsection{Equations of motion}\label{sectionnblie}

Using the above notations, the equations of motion of the $N$-body problem
are the following,
\begin{eqnarray}
\dot x_{im} & = &v_{im}, \label{nbody} \\
\dot v_{im} & = &-G\sum\limits_{j=1, j\ne i}^{K}m_j\frac{x_{im}-x_{jm}}{\rho_{ij}^3} \label{nbody2},
\end{eqnarray}
where $G$ is the Newtonian gravitational constant and $\rho_{ij}$ is
the distance between the $i$th and $j$th 
body, i.e. 
\begin{equation}
\rho_{ij}^2=\sum\limits_m(x_{jm}-x_{im})^2=A_{ijm}A_{ijm}.
\end{equation}
We also introduce the following new variables and differential operators:
\begin{eqnarray}
A_{ijm} & := & x_{im}-x_{jm}, \\
B_{ijm} & := & v_{im}-v_{jm}, \\
\Lambda_{ij} & := &  A_{ijm}B_{ijm}, \\
\phi_{ij} & := & \rho_{ij}^{-3}, \\
D_{im} & := & \frac{\partial}{\partial x_{im}}, \\
\Delta_{im} & := & \frac{\partial}{\partial v_{im}}.
\end{eqnarray}
With these notations the Lie-operator of the equations of motion can be
written as,
\begin{equation}
L_0=v_{im}D_{im} - G\sum\limits_i\left[\left(\sum\limits_{j=1, j\ne i}^{K}m_j\phi_{ij}A_{ijm}\right)\Delta_{im}\right].
\end{equation}

In \apref{appendixrecurrnbody} we prove that the recurrence relations for 
the variables $x_{im}$, $A_{ijm}$, $B_{ijm}$, $\Lambda_{ij}$,
$v_{im}$ and $\phi_{ij}$ is the following system of equations:
\begin{eqnarray}
L_0^{n+1}x_{im} & = & L_0^nv_{im}, \label{nbrec}\\
L_0^{n}A_{ijm} & = & L_0^nx_{im}-L_0^nx_{jm},  \\
L_0^{n}B_{ijm} & = & L_0^nv_{im}-L_0^nv_{jm},  \\
L_0^{n+1}v_{im} & = & -G\mathop{\sum\limits_{j=1}}\limits_{j\ne i}^{K}m_j\left[\sum\limits_{k=0}^{n}\binom{n}{k}L_0^k\phi_{ij}L_0^{n-k}A_{ijm}\right],  \\
L_0^{n}\Lambda_{ij} & = & \sum\limits_{k=0}^{n}\binom{n}{k}L_0^kA_{ijm}L_0^{n-k}B_{ijm},  \\
L_0^{n+1}\phi_{ij} & = & \rho_{ij}^{-2}\sum\limits_{k=0}^{n}F_{nk}L_0^{n-k}\phi_{ij}L_0^k\Lambda_{ij}, \label{nbrec2}
\end{eqnarray}
where $F_{nk}=(-3)\binom{n}{k}+(-2)\binom{n}{k+1}$. We note that
$F_{nk}$ is equivalent to the matrix $A_{nk}$ introduced in H\&D.

\subsection{Linearized equations}

For the linearized coordinates and velocities we introduce the
variables $\xi_{im}$ and $\eta_{im}$, respectively. Therefore, using
the Lemma, we get the Lie-derivatives of the
linearized variables, namely,
\begin{eqnarray}
L^n\xi_{im} & = & (\xi_{kp}D_{kp} + \eta_{kp}\Delta_{kp})L^nx_{im}, \label{nblinliederiv}\\
L^n\eta_{im} & = & (\xi_{kp}D_{kp} + \eta_{kp}\Delta_{kp})L^nv_{im} \label{nblinliederiv2}.
\end{eqnarray}
To obtain recurrence relations we have to introduce other auxiliary
quantities. First, we form two vectors which contain all of the linearized
variables and the differential operators:
\begin{eqnarray}
\Xi_{kp} & := & \binom{\xi_{kp}}{\eta_{kp}}, \\
\mathcal{D}_{kp} & := & \binom{D_{kp}}{\Delta_{kp}}.
\end{eqnarray}
Therefore, one can write $\Xi\cdot\mathcal{D}=\Xi_{kp}\mathcal{D}_{kp}=
\xi_{kp}D_{kp}+\eta_{kp}\Delta_{kp}$ which simplifies the notation
of the scalar products appearing in \eqrefs{nblinliederiv}{nblinliederiv2}:
\begin{eqnarray}
L^n\xi_{im} & = & \Xi\cdot\mathcal{D}L^nx_{im}, \\
L^n\eta_{im} & = & \Xi\cdot\mathcal{D}L^nv_{im}.
\end{eqnarray}
Second, let us introduce $\alpha_{ijm} := \xi_{im}-\xi_{jm}$ and
$\beta_{ijm} := \eta_{im}-\eta_{jm}$. With these newly introduced variables
and expressions we can derive the recurrence formulae for the
linearized variables. The calculations are presented in 
\apref{appendixrecurrnbodylin} in more details, and the result is
\begin{eqnarray}
L^{n+1}\xi_{im} & = & L^n\eta_{im}, \label{nblinrec} \\
L^{n}\alpha_{ijm} & = & L^n\xi_{im}-L^n\xi_{jm}, \\
L^{n}\beta_{ijm} & = & L^n\eta_{im}-L^n\eta_{jm}, \\
\Xi\cdot\mathcal{D}L^n\Lambda_{ij} & = & \sum\limits_{k=0}^{n}\binom{n}{k}\left[
L^k\alpha_{ijm}L^{n-k}B_{ijm}+\right. \nonumber \\
 & & + \left.L^kA_{ijm}L^{n-k}\beta_{ijm}\right], \\
L^{n+1}\eta_{im} & = & -G\sum\limits_{j=1, j\ne i}^{K}m_j 
\left\{\sum\limits_{k=0}^n\binom{n}{k}\left[\%_1\right]\right\}, \\
\%_1 & = & (\Xi\cdot\mathcal{D}L^k\phi_{ij})L^{n-k}A_{ijm}+ \nonumber \\
& &  + L^k\phi_{ij}L^{n-k}\alpha_{ijm} \nonumber \\
\Xi\cdot\mathcal{D}L^{n+1}\phi_{ij} & = & 
-2\rho_{ij}^{-2}\alpha_{ijm}A_{ijm}L^{n+1}\phi_{ij} + \nonumber \\
& & + \rho_{ij}^{-2}\sum\limits_{k=0}^{n}F_{nk}
\left[\%_2\right], \label{nblinrec2} \\
\%_2 & = & (\Xi\cdot\mathcal{D}L^{n-k}\phi_{ij})L^k\Lambda_{ij}+ \nonumber \\
& & + L^{n-k}\phi_{ij}(\Xi\cdot\mathcal{D}L^k\Lambda_{ij}). \nonumber
\end{eqnarray}
For the initialization of the recursion we have to calculate 
$\Xi\cdot\mathcal{D}L^0\phi_{ij}\equiv\Xi\cdot\mathcal{D}\phi_{ij}$. 
It is easy to show that 
\begin{equation}
\Xi\cdot\mathcal{D}\phi_{ij}=\Xi\cdot\mathcal{D}\rho_{ij}^{-3}=
-3\rho_{ij}^{-5}\alpha_{ijm}A_{ijm}. \label{xidphi}
\end{equation}

We have some remarks concering the derivation and evaluation of 
the above formulae. 
First, we did not need the linearized equations explicitly to derive
the recurrence relations for the linearized variables. 
Second, because of the symmetry properties of the variables, we do not have
to calculate all of the matrix elements: we know that
the tensors $A_{ijm}$, $B_{ijm}$, $\alpha_{ijm}$, $\beta_{ijm}$ are
antisymmetric for swapping the indices $i$ and $j$ and the 
matrices $\Lambda_{ij}$, $\phi_{ij}$, $\Xi\cdot\mathcal{D}\Lambda_{ij}$ and 
$\Xi\cdot\mathcal{D}\phi_{ij}$ are symmetric.
Because distances are defined only between different bodies, the
diagonal matrix elements of $\rho_{ii}$ and their derived ($\phi_{ii}$,  %nem tudom itt a derived mit takar, de nem jobb a derivatives ?
$L^n\rho_{ii}$, $L^n\phi_{ii}$, $\Xi\cdot\mathcal{D}L^n\rho_{ii}$, \dots)
are not defined. 

\subsection{Motion in the reference frame of one of the bodies}

In the description of the Solar System or in perturbation theory,
the equations of motion are transformed 
into a reference frame whose origin coincides with one of the bodies.
Practically, it is the body with the largest mass, in
the Solar System it is the Sun (where all
orbital elements are defined relatively to the Sun). Therefore, it could
prove useful to have the recurrence relations both for the equations of 
motion and for the linearized part of the equations in this reference frame.

Let us define the central body as the body with the index of $i=0$.
Altogether we have $1+K$ bodies, where the other ones are indexed
by $i=1,\dots,K$.
For simplicity, denote its mass by $\mathcal{M}\equiv m_0$. In an intertial frame,
the equations of motion can be splitted into two parts, namely
\begin{eqnarray}
\dot x_{im} & = & v_{im}, \label{submassi}\\
\dot v_{im} & = & - G\sum\limits_{j=0, j\ne i}^{K} m_j\rho_{ij}^{-3}(x_{im}-x_{jm}), \\
\dot x_{0m} & = & v_{0m}, \label{submass0}\\
\dot v_{0m} & = & - G\sum\limits_{j=1}^{K} m_j\rho_{0j}^{-3}(x_{0m}-x_{jm}). 
\end{eqnarray}
Following the usual steps, the equations of motion in the fixed frame
can easily be derived by subtracting \eqref{submass0} from the equations
of~(\ref{submassi}) for all $i$ indices. Let us define the new variables
\begin{eqnarray}
r_{im} & := & x_{im} - x_{0m}, \\
w_{im} & := & v_{im} - v_{0m}, \nonumber \\
\rho_i & := & \rho_{0i} = \rho_{i0}, \\
\phi_i & := & \rho_i^{-3}. \nonumber
\end{eqnarray}
Note that the quantities $\rho_i$ and $\rho_{ij}$, like so 
$\phi_i$ and $\phi_{ij}$ are distinguished only by the number of their indices.
Obviously, $A_{ijm}=x_{im}-x_{jm}=r_{im}-r_{jm}$ and 
$B_{ijm}=v_{im}-v_{jm}=w_{im}-w_{jm}$. Thus using the relative (non-inertial)
coordinates and velocities, the equations of motion
in more compact form are
\begin{eqnarray}
\dot r_{im} & = & w_{im}, \label{fixednb}\\
\dot w_{im} & = & -G(\mathcal{M}+m_i)\phi_i r_{im} - \nonumber \\
	& & - G\sum\limits_{j=1, j\ne i}^{K}m_j\left[\phi_{ij}A_{ijm}+\phi_j r_{jm}\right]. \label{fixednb2}
\end{eqnarray}
Without going into details, we preset the recurrence relations of
the Lie-derivatives, including the linearized variables 
in \apref{appendixfixednb}. Some speed-up considerations with which 
the required number of operations can definitely be decreased are
presented in \apref{appendixspeedup}.

%%%%%%%%%%%%%%%%%%%%%%%%%%%%%%%%%%%%%%%%%%%%%%%%%%%%%%%%%%%%%%%%%%%%%%%%%%%%%%

\begin{table}
\caption{Timing data for the Runge-Kutta integrators, the Bulirsch-Stoer 
integrator and for the Lie-integrator
for different orders. See text for further details.}\label{table:timing}.
\begin{tabular}{lllll}
\hline	
Integrator		& CPU	& \multicolumn{3}{l}{Stepsize,
				$\Delta t_{(\varepsilon)}^{[\mathrm{method}]}$
				for $\varepsilon=$}		\\
$\mathrm{[method]}$	& time	& $2.4\cdot10^{-11}$	
				& $2.4\cdot10^{-12}$
				& $2.4\cdot10^{-13}$		\\
\hline
RK4			& 0.302	& 0.0140 & 0.0091 & 0.0026	\\
RKN5/6/			& 0.460	& 0.0243 & 0.0153 & 0.0097	\\
RKN7/8/			& 0.941	& 0.2521 & 0.1962 & 0.1548	\\
BS			& 8.578 & 2.0172 & 1.7734 & 1.5773	\\
$M=6$			& 0.916	& 0.0829 & 0.0603 & 0.0437	\\
$M=7$			& 1.169	& 0.1095 & 0.0796 & 0.0556	\\
$M=8$			& 1.421	& 0.2271 & 0.1776 & 0.1339	\\
$M=9$			& 1.706	& 0.2916 & 0.2275 & 0.1792	\\
$M=10$			& 2.004	& 0.4332 & 0.3541 & 0.2912	\\
$M=11$			& 2.336	& 0.5535 & 0.4480 & 0.3525	\\
$M=12$			& 2.689	& 0.6941 & 0.5762 & 0.4715	\\
$M=13$			& 3.048	& 0.9055 & 0.7770 & 0.6137	\\
$M=14$			& 3.411	& 0.9352 & 0.8555 & 0.7113	\\
$M=15$			& 3.810	& 0.9414 & 0.8805 & 0.8789	\\
$M=16$			& 4.223	& 0.9492 & 0.9414 & 0.9383	\\
\hline
\end{tabular}
\end{table}

%%%%%%%%%%%%%%%%%%%%%%%%%%%%%%%%%%%%%%%%%%%%%%%%%%%%%%%%%%%%%%%%%%%%%%%%%%%%%%

\section{Performance and comparisons}

\begin{figure*}
\resizebox{8cm}{!}{\includegraphics{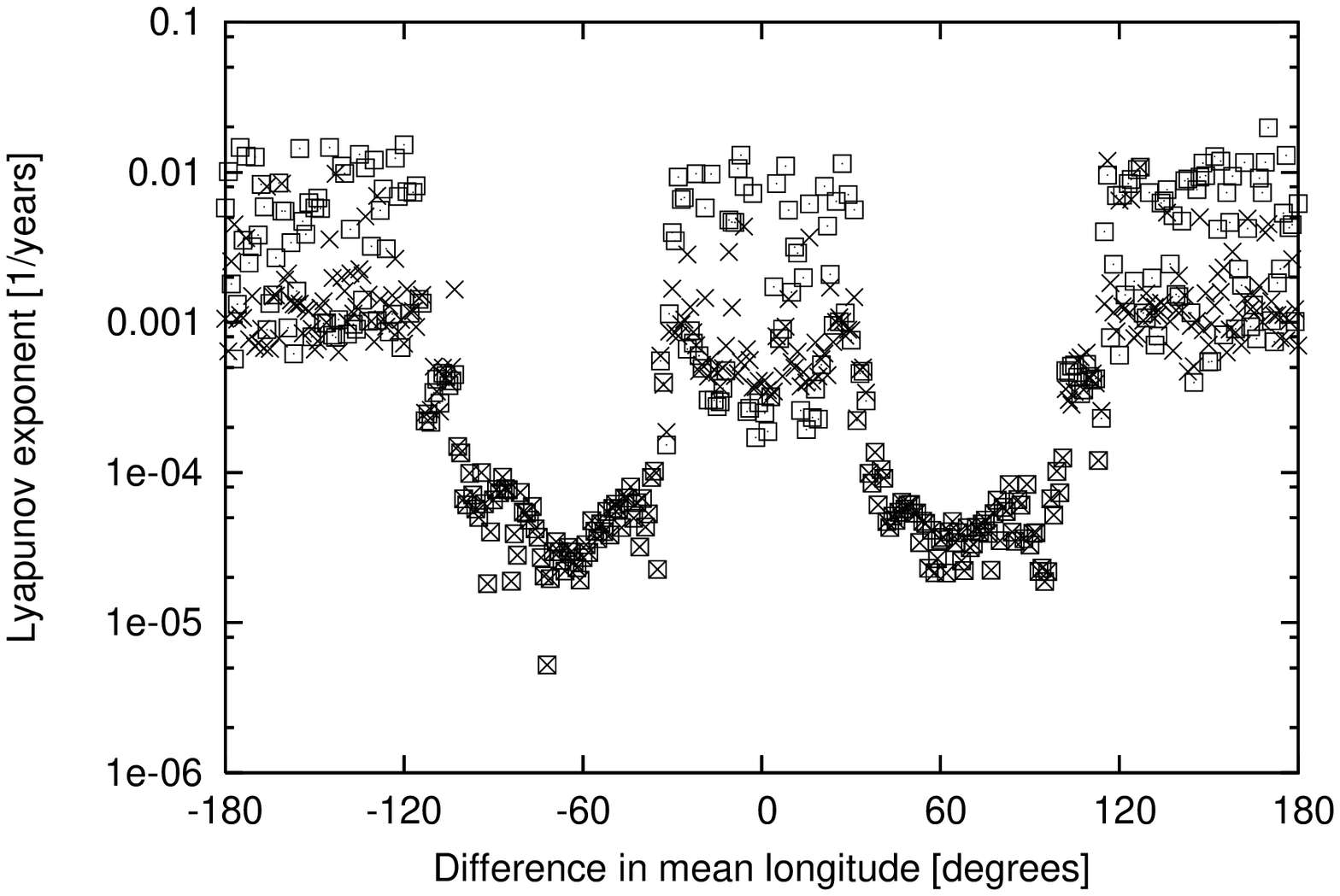}}%
\resizebox{8cm}{!}{\includegraphics{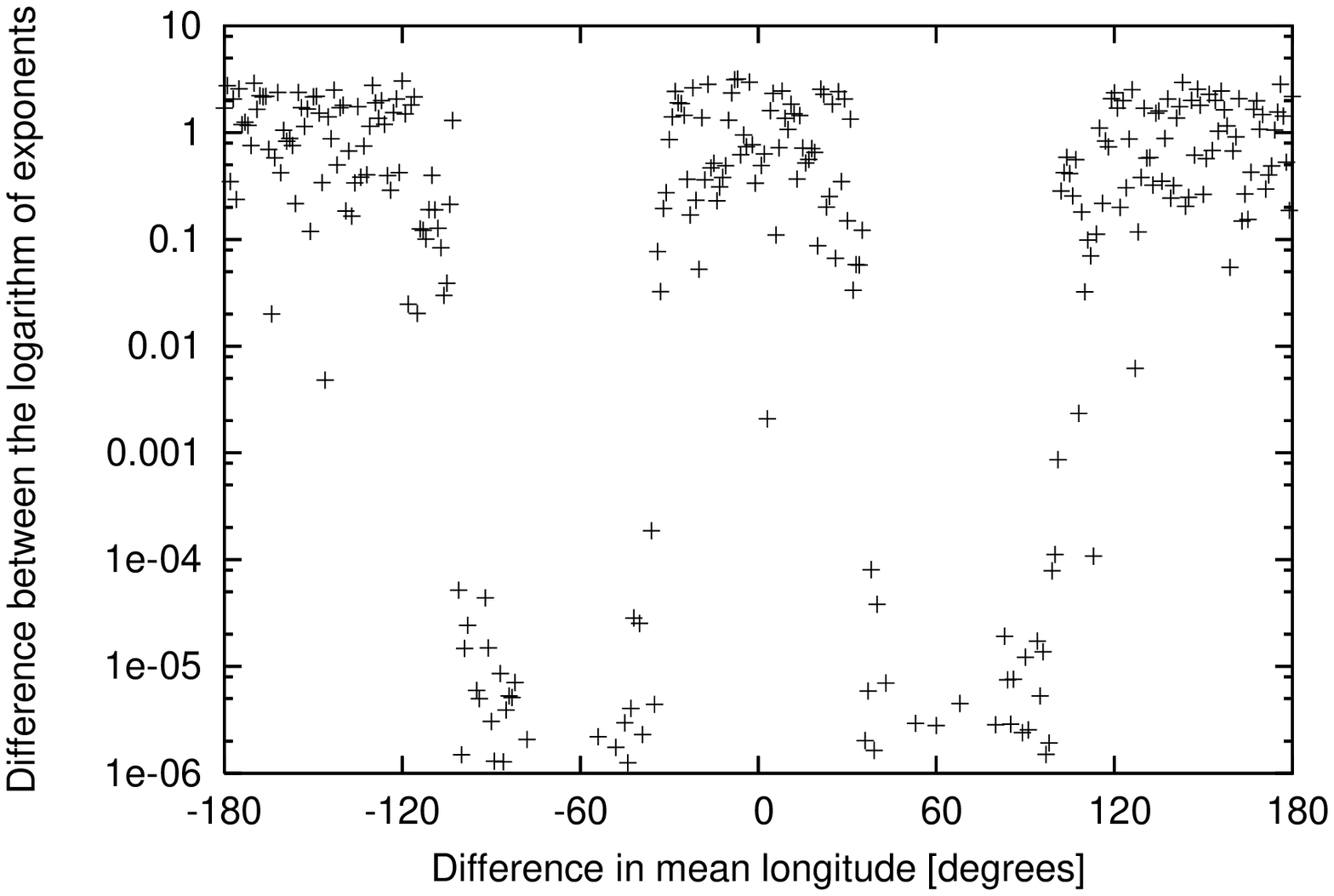}}
\caption{%
The LCIs for a fictious asteroid having the same orbit as 
Jupiter. In the left panel, one can see the derived indicators by
the method of RKN7/8/ (crosses) and using Lie-integration (empty squares) as
the function of $\Delta\lambda$. In
the right panel, the ratios $\chi(\Delta\lambda)$ 
defined by \eqref{lyapratio} are plotted.}
\label{fig:lyaps}
\end{figure*}

We have implemented the method of Lie-integration as a standalone program,
written in ANSI C, with the following capabilities. The program is able
to integrate the equations of motion of the $N$-body problem in
the reference frame of one of the bodies (see \eqrefs{fixednb}{fixednb2}) and
parallelly, the program approximates the LCE by the Lyapunov Characteristic
Indicator (LCI) of the
system using the solution of the linearized equations. For the method
of integration one could use the classical fourth order Runge-Kutta 
\citep[see][]{press1992} and Runge-Kutta-Nystrom integrators 
\citep[namely, RKN5/6/ and RKN7/8/, see][]{fehlberg1972,dormand1978},
the Bulirsch-Stoer integrator \citep[BS, see also][]{press1992}
as well the Lie-integration method (see \eqrefs{fixnbrec}{fixnbrec2} and 
\eqrefs{fixnblinrec}{fixnblinrec2} in \apref{appendixfixednb}), up to arbitrary order $M$. 
The program is also
able to figure out the optimal stepsizes to satisfy a pre-defined
accuracy. The accuracy is derived using the differences in the
mean longitude which is the fastest changing orbital element.
This type of accuracy control can be found in many integrators (e.g.~ORBIT9)
where the dimensionless accuracy is defined as the difference of the
mean longitudes between the exact and approximated solution 
(in radians) divided by the the square of the number
of revolutions, namely
\begin{equation}
\varepsilon=
\frac{|\Delta\lambda|(\mathrm{radians})}{N^2_{\mathrm{revolution}}}\label{accdef}
\end{equation}
\citep[see][for a more detailed explanation]{milani1988}.

\begin{figure*}
\resizebox{8cm}{!}{\includegraphics{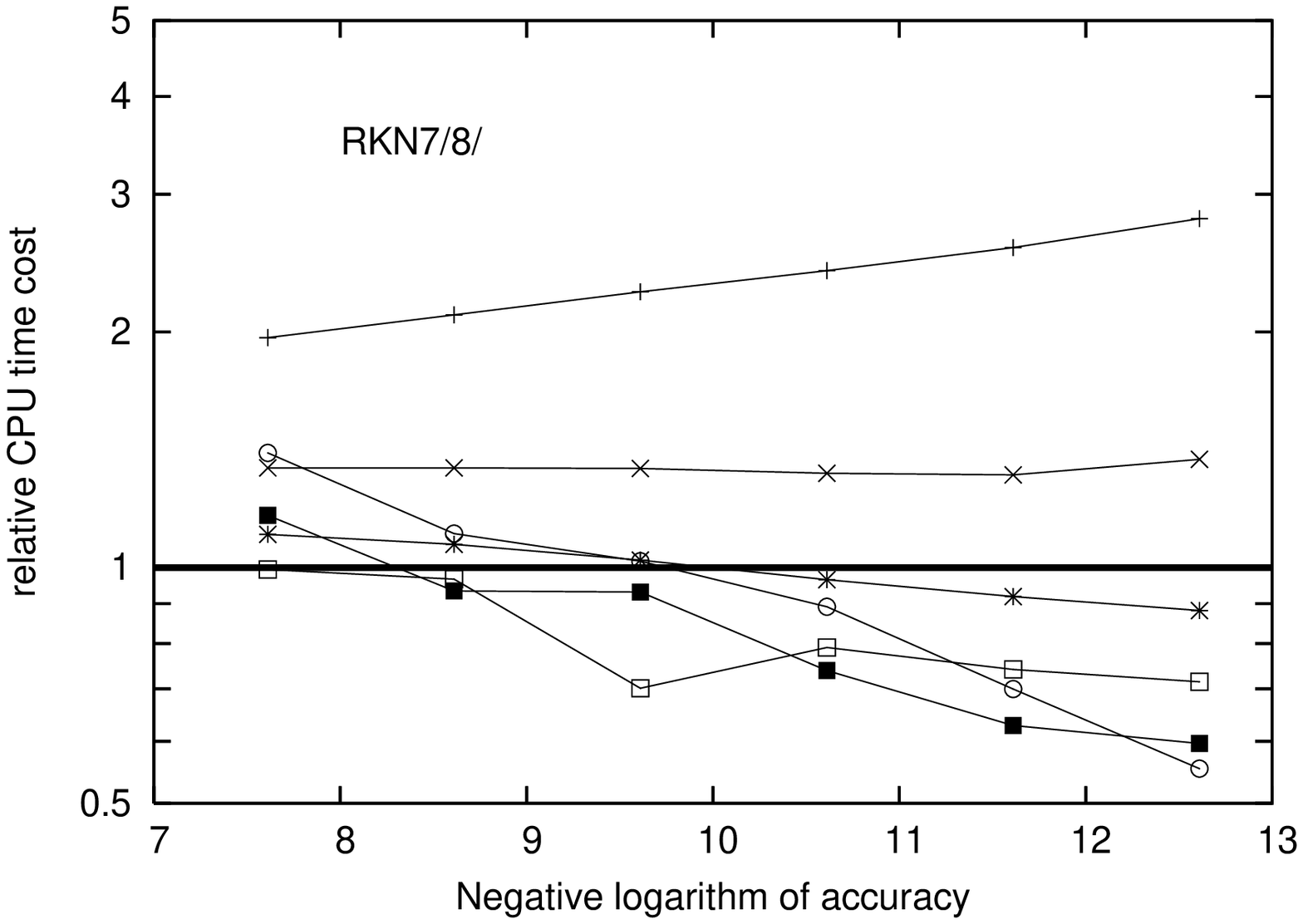}}%
\resizebox{8cm}{!}{\includegraphics{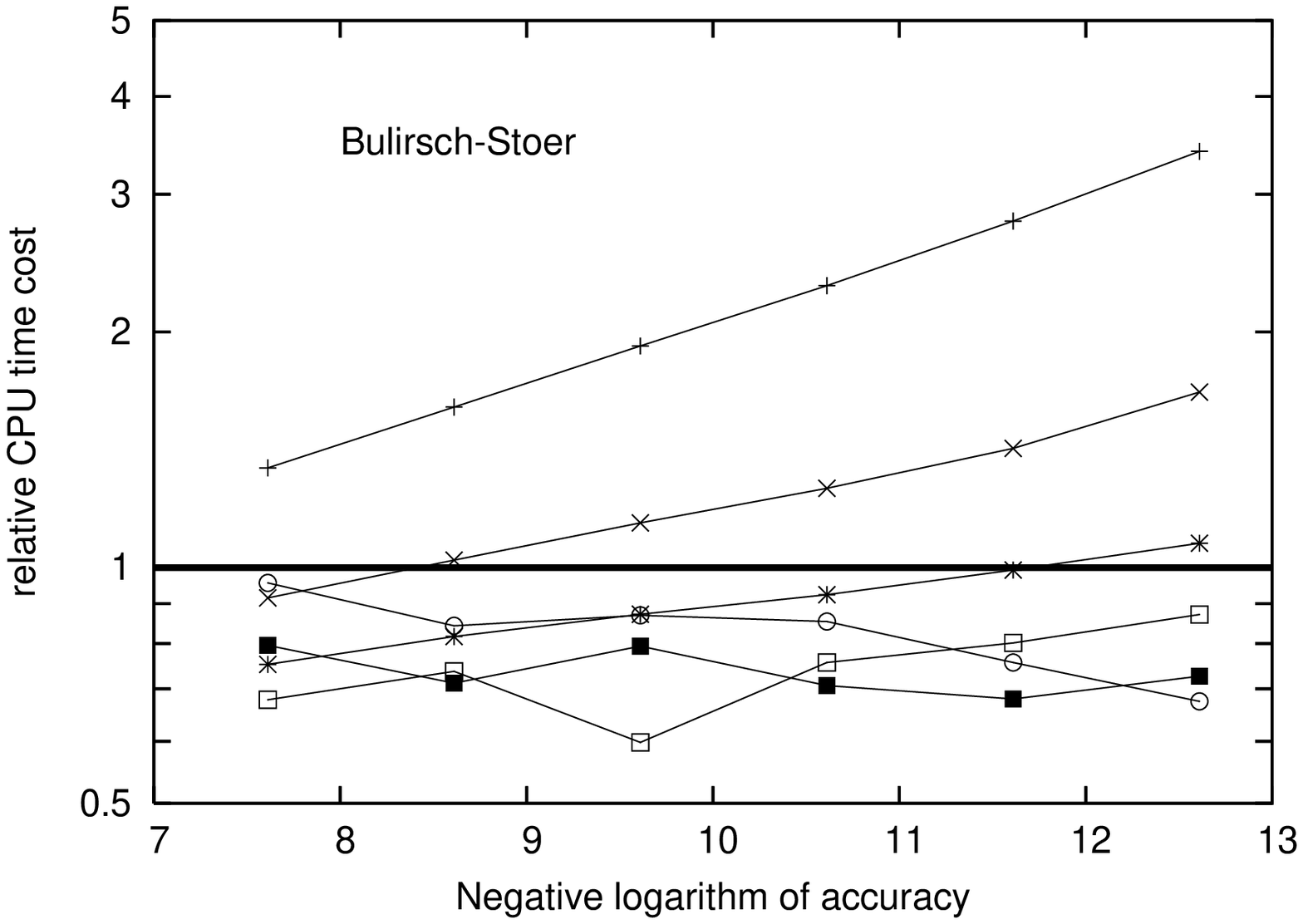}}
\caption{%
The relative cost in the CPU time as the function of the desired precision for
the Lie-integration against the RKN7/8/ (left panel) and the 
Bulirsch-Stoer method (right panel) for the model system of 
Sun -- Jupiter -- Saturn.
The curves show the cost for different orders of the Lie-integration
(plus signs for $M=6$, crosses: $M=8$, stars: $M=10$, open squares: $M=12$,
filled squares: $M=14$ and open circles $M=16$). The thick solid
line shows the unity cost, where the RKN/BS and Lie-integration methods have 
the same performance. Smaller costs represent lower CPU usage, 
therefore higher gain and better performance.}\label{fig:costnone}
\end{figure*}

As an initial test, we have compared the LCIs
computed by two different integration methods, namely RKN7/8/ and
the Lie-integration with the order of $M=8$. The dynamical system
is the spatial Sun -- Jupiter -- Saturn -- test particle spatial 
restricted four-body system 
where the latter has the same orbit as the Jupiter has. The
LCIs are calculated as the function of the difference in the
mean longitudes of Jupiter ($\lambda_{\mathrm{J}}$)
and the test particle ($\lambda_{\mathrm{m}}$) while all other 5
initial orbital elements are equal to those of Jupiter. The 
results are plotted in Fig.~\ref{fig:lyaps}. In the left panel
of Fig.~\ref{fig:lyaps}, one can see the derived indicators by
the method of RKN7/8/, $\mathrm{LCI}_{\mathrm{RKN7/8/}}$ and using 
Lie-integration, $\mathrm{LCI}_{\mathrm{Lie}}$ as
the function of $\Delta\lambda=\lambda_{\mathrm{m}}-\lambda_{\mathrm{J}}$. In
the right panel, the absolute value of the base-10 logarithm of the ratio
of the indicators, namely
\begin{equation}
\chi=\left|\log_{10}\left(\frac{\mathrm{LCI}_{\mathrm{RKN7/8/}}}{\mathrm{LCI}_{\mathrm{Lie}}}\right)\right|,\label{lyapratio}
\end{equation}
are plotted resulted by this two integration method. 
Note that the integration length is $10^6~\mathrm{yrs}$, therefore 
the LCIs concerning to regular solutions are saturated around 
$\approx 10^{-5} - 10^{-6}~1/\mathrm{yr}$.
It can easily be seen that in the stable regions (around the two Lagrangian
points at $\Delta\lambda=-60^\circ$ and $\Delta\lambda=+60^\circ$) the results
of the two methods are very similar, the magnitude of the differences
between them is $\approx 10^{-5}$.
In the chaotic regions, the two methods yielded different LCIs but
their magnitudes were always the same. 

\begin{figure*}
\resizebox{8cm}{!}{\includegraphics{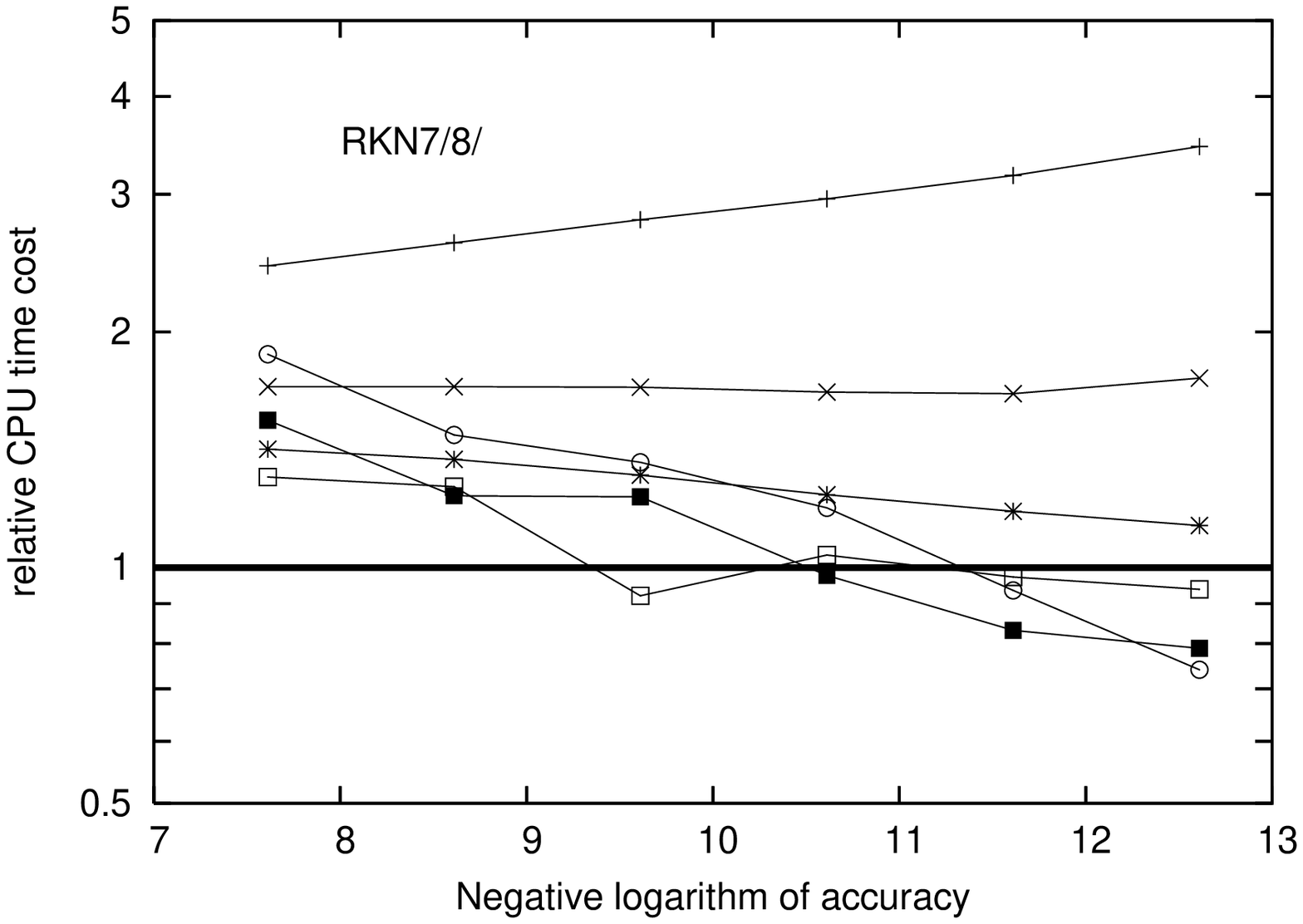}}%
\resizebox{8cm}{!}{\includegraphics{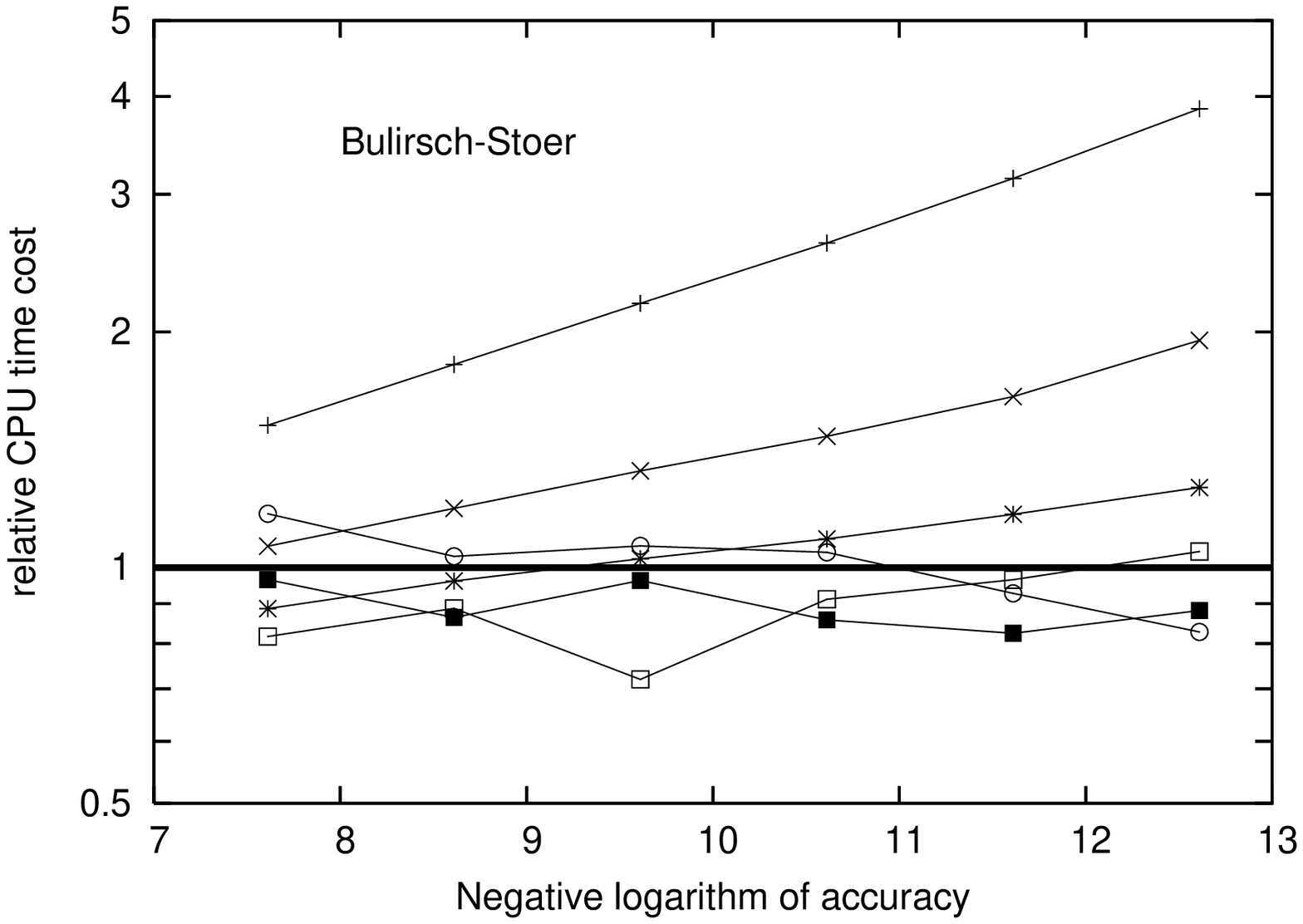}}
\caption{%
The relative cost in the CPU time as the function of the desired accuracy for
the Lie-integration against the RKN7/8/ (left panel) and the Bulirsch-Stoer method
(right panel) for the model system of Sun -- Jupiter -- Saturn -- fictitious
asteroid,
while the linearized equations are also evaluated only for the
massless test particle.
The curves show the cost for different orders of the Lie-integration
(see also Fig.~\ref{fig:costnone}).}\label{fig:costpart}
\end{figure*}

\subsection{Performance analysis}

We have compared the efficiency of the Lie-integrator and the other implemented
integrators. Here we give how much CPU time is required to
integrate the equations of motion with RK4, RKN5/6/, RKN7/8/, BS and with
the Lie-integration and parallelly the linearized equations to
get the result with a previously given accuracy. The ratio of the net CPU
times is the relative cost:
\begin{equation}
\mathrm{cost}^\mathrm{[other]} := \frac{\tau_{\mathrm{CPU}}^{\mathrm{[Lie]}}}{\tau_{\mathrm{CPU}}^{\mathrm{[other]}}}.
\end{equation}
As one can see, the smaller the cost is the more efficient the 
Lie-integration is. It should be kept in mind that this relative cost 
does not only depend on the other method but also on the order of the Lie-integration
and the desired accuracy. Going into the details, the cost has been measured
indirectly by the following way. It can be said that any of the integration
algorithms, the RK-based ones, the BS and the Lie-integration use the
same CPU time \emph{per step} independently from the stepsize%
\footnote{In our tests, in the BS method the adaptive variation of 
the number of MMID substeps has been disabled, i.e.~the extrapolation
is performed after the same sequence of number of substeps: it yields
an evaluation time which is independent from the stepsize.}.
Let us denote this atomic CPU time by $\tau_{(0)}^{[\mathrm{method}]}$.
Therefore,
if the optimal stepsize $\Delta t_{(\varepsilon)}^{[\mathrm{method}]}$ is 
known for a given method and accuracy, 
the total CPU time can easily be calculated:
\begin{equation}
\tau_{\mathrm{CPU}}^{[\mathrm{method}]}=\tau_{(0)}^{[\mathrm{method}]}\frac{T}{\Delta t_{(\varepsilon)}^{[\mathrm{method}]}},
\end{equation}
where $T$ is the total length of the integration. Because the relative cost is
the ratio of two such value of $\tau_{\mathrm{CPU}}$ for two methods,
the total length of the integration cancels. The atomic CPU time 
can easily be measured, the only unknown is the $\Delta t$ optimal stepsize
for the different methods. The latter is determined by the following way. The
exact mean longitude for the fastest rotating planet is derived for a given 
time-span (which is defined by the accuracy, see \eqref{accdef}) with
an appropriately small stepsize. After it, the stepsize is increased
iteratively by a bracketing algorithm until the integration yields
a mean longitude which differs from the exact one by the $\Delta\lambda$
value determined also by \eqref{accdef}. We should note that this implies 
that the stepsize $\Delta t$ is constant during the integration.
In Table~\ref{table:timing} we summarize
these timing values for some values of accuracy and for the Runge-Kutta
methods, for the Bulirsch-Stoer method as well as for the Lie-integration 
method for orders $M=6,\dots,16$
for the dynamical system of Sun -- Jupiter -- Saturn -- test particle 
extended with the linearized equations of the latter.
The second column contains the atomic CPU time\footnote{measured on 
an Athlon XP 1800+ processor with GCCv4.1.2 compiler, in the units of
$10^{-6}$ seconds.}, while the other three columns show the optimal 
stepsize $\Delta t_{(\varepsilon)}^{[\mathrm{method}]}$, derived by
the above manner for the accuracies $\varepsilon=2.4\cdot10^{-11}$,
$2.4\cdot10^{-12}$ and $2.4\cdot10^{-13}$ respectively.
Thus, the cost can be derived by the fractions of the 
appropriate values taken from this table, namely: 
\begin{equation}
\mathrm{cost}^\mathrm{[m1] against [m2]}=
\frac	{\tau_{(0)}^{[\mathrm{m1}]}\Delta t_{(\varepsilon)}^{[\mathrm{m2}]}}
	{\tau_{(0)}^{[\mathrm{m2}]}\Delta t_{(\varepsilon)}^{[\mathrm{m1}]}},\label{costdef}
\end{equation}
where $\mathrm{[m1]}$ and $\mathrm{[m2]}$ index the two methods to be 
compared. We should note that timing values were not only derived
for these values of accuracy as it can be read from Table~\ref{table:timing}
and we have made timing measurements when the linearized equations are
omitted. See next sections for more details and for other plots.

\subsection{Efficiency as the function of the accuracy}

In Fig.~\ref{fig:costnone} the relative cost of the Lie-integration
against the RKN7/8/ and the Bulirsch-Stoer integration method are plotted 
for the three-body problem of Sun -- Jupiter -- Saturn as the function of 
the accuracy. Different curves show the cost for different orders of the 
Lie-integration between $6$ and $16$. It can easily be seen that for higher 
orders and below a critical accuracy the Lie-integration is more efficient 
than RKN7/8/ and for higher orders, the Lie-integration is more efficient
than the Bulirsch-Stoer method almost independently from the accuracy.
Note that in this plot the linearized equations are omitted from the
calculations.

In Fig.~\ref{fig:costpart} we have plotted the cost of the Lie-integration
against the methods as above but the dynamical system is extended
with a massless test particle and for the latter the linearized equations
are also solved. The qualitative behaviour of the cost as the function
of the accuracy and the orders of the Lie-series is almost the
same as in Fig.~\ref{fig:costnone}. We note that the different
methods used in the RKN, BS and Lie-integration to evaluate 
the linearized equations result different number of operations, therefore
the costs won't be exactly the same. Namely, the relative CPU time cost
of the Lie-integration against the other methods is slightly larger when the 
linearized equations are solved parallely.

As a conclusion, we can say that omitting the linearized part orders 
below $M\approx 10$ the Lie-integration method is inferior to
the RKN7/8/, while the equations are extended with the linearized
equations for the massless particle, the Lie-integration is more
effective than RKN7/8/ for orders larger than $M\approx12$ below
a certain accuracy about $\varepsilon\approx10^{-11}$. Comparing with
the BS method, the Lie-integration is more effective for orders larger
than $M\approx 8$ and $M\approx 10$ when the linearized equations are
omitted or not, almost independently from the accuracy. We note that
the Lie-integration is effective with more than a magnitude (or more) than the
lower-order Runge-Kutta methods, as it can easily be derived from 
Table.~\ref{table:timing} and \eqref{costdef}.

\begin{figure}
\resizebox{8cm}{!}{\includegraphics{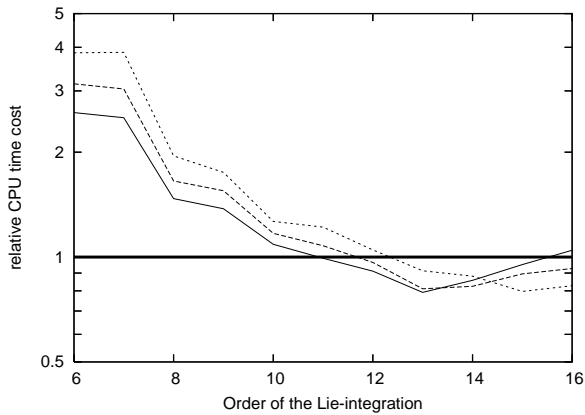}}
\caption{%
The relative cost in the CPU time as the function of the order of the Lie-integration
against the RKN7/8/ method for the model system of 
Sun -- Jupiter -- Saturn -- fictitious asteroid. In these runs the 
linearized equations are also evaluated for the massless test particle.
The thin solid line, the long dashed line and the dashed line show the cost
for the accuracy of $2.4\cdot10^{-11}$, $2.4\cdot10^{-12}$ and
$2.4\cdot10^{-13}$, respectively. The thick line marks the unity
cost, below which the Lie-integration is more efficient.}\label{fig:ordercost}
\end{figure}

\subsection{Efficiency as the function of the order}

As it was written in the introduction, the method of Lie-integration
approximates the Taylor-expansion of the solution up to a finite order.
One can easily prove that the appropriate order, $n$ of a Taylor-series to 
obtain a certain accuracy of a periodic
function defined on an interval is proportional to the length, $L$ of this 
interval. The concept of the proof is as follows. 
An adequately smooth periodic function can be approximated
as a sum of sine (and cosine) functions, the so-called Fourier terms. 
The sine function, $\sin(x)$ can be expanded as
\begin{equation}
\sin(x)=\sum\limits_{k=0}^\infty (-1)^k\frac{x^{2k+1}}{(2k+1)!}.
\end{equation}
To obtain an accuracy of unity, the last ($n=2k+1$th) term of the series should
be the solution of $x^n\equiv L^n\approx n!$. Therefore, using Stirling's 
approximation, one gets 
\begin{equation}
n\log L = \log(n!)\approx n\log n - n,
\end{equation}
so $n\approx\mathrm{e}\cdot L$, which means $n\sim L$. This is true for
all Fourier terms of the expansion of a periodic function.

Thus one can assume that to obtain a certain accuracy, 
the $M$ number of the terms in the Lie-series is proportional to the length
of the integration stepsize, namely $\Delta t\approx\kappa M$. 
The total number of arithmetical
operations, therefore the required CPU time is a quadratic function of the
order $M$: $\tau_{\mathrm{CPU}}=\mathrm{Ordo}(M^2)$. To be more precise, the 
required CPU time is $\tau_{\mathrm{CPU}}=\alpha+\beta M+\gamma M^2$, 
for smaller $M$'s, the first two terms, $\alpha$ and $\beta M$ are not 
negligible. Therefore,
to integrate the equations over an interval $T$ requires 
\begin{equation}
\tau_{\mathrm{CPU}}^{\mathrm{total}}\approx(\alpha+\beta M+\gamma M^2)\frac{T}{\Delta t}=
(\alpha+\beta M+\gamma M^2)\frac{T}{\kappa M}\label{costfunctorder}
\end{equation}
CPU time, which, depending on the ratios of the constants $\alpha$, $\beta$
and $\gamma$ has a minimum corresponding to the optimal order of the 
Lie-integration method.

We have tested this type of dependency of the CPU time on the order of
the Lie-integration. The results are plotted in Fig.~\ref{fig:ordercost},
for three values of accuracy, while the dynamical system is 
the restricted four body problem of 
Sun -- Jupiter -- Saturn -- test particle extended with the linearized
equations respecting to the latter. As it was assumed above, 
the relative cost has a minimum corresponding to the optimal order
and the value of this minimum is around $M\approx13-15$, depending
on the accuracy. It can also be seen that for larger values of $M$
the cost increases which means worse efficiency, as it is 
expected from \eqref{costfunctorder}.
What is more 
interesting that the position of the minimum clearly 
depends on the accuracy: the better the accuracy is the larger the 
optimal value of $M$ is. This might imply another kind of adaptive integration
method where not only the stepsize varies but the order of 
the Lie-integration.

\subsection{Implemetaion of the method}

We have implemented the method of Lie-integration of the $N$-body problem
as a standalone ANSI C program, extended with the linearized equations 
and the capability to calculate the LCIs. The version of the program
which can integrate the motion of 1+3 bodies and was used in our
benchmarks can be downloaded from the address \texttt{http://cm.elte.hu/lie}
as a single \texttt{.tar.gz} archive. The full version which is capable to
integrate the motion of arbitrary number of bodies can be requested from the
first author via e-mail. All versions of this code are designed to work
on UNIX-like environments.

%%%%%%%%%%%%%%%%%%%%%%%%%%%%%%%%%%%%%%%%%%%%%%%%%%%%%%%%%%%%%%%%%%%%%%%%%%%%%%

\section{Discussion and summary}

In this paper we have presented a lemma with which 
recurrence relations can be derived for the Lie-integration of 
linearized equations. We have
demonstrated the usage of this lemma on the H\'enon-Heiles system, and
thereafter applied it to the equations of the $N$-body problem, including
the non-inertial equations where the origin of the reference frame
is fixed to one of the bodies. Our performance 
comparisons have shown that although these recurrence formulae 
are rather complicated, they can efficiently be used for integrations 
where high accuracy is required. We have investigated realistic dynamical 
systems for these comparisons:
using the lemma, the recurrence relations were determined and using 
the Lie-integration technique, the LCIs for a fictitious asteroid was computed.
The method of LCIs is the basis for many modern chaos
detection methods, therefore our lemma and the derived Lie-integration
method can widely be used in various kind of dynamical 
investigations, providing a faster alternative to the currently used
techniques. We have checked the
efficiency as the function of the order and accuracy. These tests have
shown that the Lie-integration is definitely more effective than
the classical RK4 and RKN5/6/ integration method and above a certain 
accuracy about $\varepsilon\approx10^{-11}$, the Lie-integration is 
more effective than the method of RKN7/8/ for orders larger than $M=10$ 
or $M=8$, whether the linearized equations are evaluated parallelly or not.
We found that the Lie-integration is more effective than the BS integrator
for orders larger than $M\approx10$, almost independently from the accuracy.

Further studies are already ongoing concerning this problem. First,
there could be several possibilities for optimization in the 
actual implementation of the Lie-integration: we expect that 
the re-ordering of the 
highly nested loops and/or the introduction of new auxiliary variables 
yield better performance. Second, some aspects of the Lie-integration
should better be analysed and understood, including the long-term 
error propagation in higher orders which is the basis of the adaptive 
extensions of this integration method. And last, we are going to develop
a more general code which is not only capable of the calculation
of LCIs but can be extended with other chaos indicators.

%%%%%%%%%%%%%%%%%%%%%%%%%%%%%%%%%%%%%%%%%%%%%%%%%%%%%%%%%%%%%%%%%%%%%%%%%%%%%%

{}

%%%%%%%%%%%%%%%%%%%%%%%%%%%%%%%%%%%%%%%%%%%%%%%%%%%%%%%%%%%%%%%%%%%%%%%%%%%%%%

\onecolumn
\appendix

% At kellene nezni a term 1, term 2, ...-ne'l az uj indexek bevezetesenel a cuccokat, hogy jol van-e gondolva...
% trivialis, de ide is be kell irni az olyan relaciokat mint L^{n+1}x = L^{n}v (A-ra, B-re is, ...), azokat, amikre _hivatkozas_van_ a fo"-fejezetben...!

\section{Recurrence relations for the $N$-body problem}
\label{appendixrecurrnbody}

Using the notations defined in \scref{sectionnblie}, we derive the 
recurrence relations for the equations of motion for the $N$-body problem.

As we have shown in \scref{sectionnblie} the Lie-operator of the
equations of motion is
\begin{equation}
L_0=v_{im}D_{im} - G\sum\limits_i\left[\left(\sum\limits_{j=1, j\ne i}^{K}m_j\phi_{ij}A_{ijm}\right)\Delta_{im}\right].
\end{equation}
This implies that the first Lie-derivatives of the coordinates and velocities
are the right-hand side of the equations of motion, namely
\begin{eqnarray}
L_0x_{im} & = & v_{im}, \label{nbodylienative} \\
L_0v_{im} & = &-G\sum\limits_{j=1, j\ne i}^{K}m_j\rho_{ij}^{-3}(x_{jm}-x_{im})=-G\sum\limits_{j=1, j\ne i}^{K}m_j\phi_{ij}A_{ijm} \nonumber.
\end{eqnarray}
The distance $\rho_{ij}$ does not depend on the velocities, like so $\phi_{ij}$,
therefore their Lie-derivatives can easily be calculated:
\begin{equation}
L_0\rho_{ij} = \sum\limits_k v_{km}D_{km}\rho_{ij}\label{nblierho}. 
\end{equation}
Therefore,
\begin{eqnarray}
L_0\rho_{ij} & = & \sum\limits_k v_{km}\frac{\partial}{\partial x_{km}}
\sqrt{\sum\limits_m(x_{jm}-x_{im})(x_{jm}-x_{im})}=
\sum\limits_k v_{km}\frac{\frac{\partial}{\partial x_{km}}\left(\sum\limits_m(x_{jm}-x_{im})(x_{jm}-x_{im})\right)}{2\sqrt{\sum\limits_m(x_{jm}-x_{im})(x_{jm}-x_{im})}}= \nonumber \\
& = & \sum\limits_k v_{km}\left[\frac{1}{2\rho_{ij}}\left(2(x_{jm}-x_{im})(\delta_{kj}-\delta_{ki})\right)\right]=
\sum\limits_k v_{km}\rho_{ij}^{-1}\left[(x_{jm}-x_{im})(\delta_{kj}-\delta_{ki})\right]= \nonumber \\
& = & \rho_{ij}^{-1}\left[(v_{jm}-v_{im})(x_{jm}-x_{im})\right]=\rho_{ij}^{-1}B_{jim}
A_{jim}=\rho_{ij}^{-1}\Lambda_{ji}=\rho_{ij}^{-1}\Lambda_{ij}.
\end{eqnarray}
Now one can calculate the Lie-derivative of $\phi_{ij}=\rho_{ij}^{-3}$:
\begin{equation}
L_0\phi_{ij}=L_0(\rho_{ij}^{-3})=-3\rho_{ij}^{-4}L_0\rho_{ij}=-3\rho_{ij}^{-5}
\Lambda_{ij}=\rho_{ij}^{-2}(-3\phi_{ij}\Lambda_{ij}).\label{lierho}
\end{equation}
With mathematical induction, one can prove that 
\begin{equation}
L_0^{n+1}\phi_{ij}=\rho_{ij}^{-2}\sum\limits_{k=0}^{n}\left[-3\binom{n}{k}-2\binom{n}{k+1}\right]L_0^{n-k}\phi_{ij}L_0^k\Lambda_{ij}. \label{liephir0}
\end{equation}
For $n=0$ this equation is equivalent to \eqref{lierho}. Let us assume that
this is true for all $m\le n$, and calculate $L_0^{n+2}\phi_{ij}$:
\begin{eqnarray}
L_0^{n+2}\phi_{ij} & = & L_0\left(\rho_{ij}^{-2}\sum\limits_{k=0}^{n}\left[-3\binom{n}{k}-2\binom{n}{k+1}\right]L_0^{n-k}\phi_{ij}L_0^k\Lambda_{ij}\right)=\nonumber \\
& = & \underbrace{L_0\left(\rho_{ij}^{-2}\right)\cdot\left(\rho_{ij}^2L_0^{n+1}\phi_{ij}\right)}_{\mathrm{Term~1}}+
      \underbrace{\rho_{ij}^{-2}\sum\limits_{k=0}^{n}\left[-3\binom{n}{k}-2\binom{n}{k+1}\right]L_0^{n+1-k}\phi_{ij}L_0^{k}\Lambda_{ij}}_{\mathrm{Term~2}}+ \nonumber \\
& & +\underbrace{\rho_{ij}^{-2}\sum\limits_{k=0}^{n}\left[-3\binom{n}{k}-2\binom{n}{k+1}\right]L_0^{n-k}\phi_{ij}L_0^{k+1}\Lambda_{ij}}_{\mathrm{Term~3}}=\%. \label{liephir1}
\end{eqnarray}
The first term is
\begin{equation}
L_0\left(\rho_{ij}^{-2}\right)\cdot\left(\rho_{ij}^2L_0^{n+1}\phi_{ij}\right)=
-2\rho_{ij}^{-1}\Lambda_{ij}\phi_{ij}\rho_{ij}^2L_0^{n+1}\phi_{ij}=
\rho_{ij}^{-2}\left(-2L_0^{n+1}\phi_{ij}\Lambda_{ij}\right). \label{liephir2}
\end{equation}
We can increase the upper limit of the first summation of term~2 
in \eqref{liephir1} from $n$ to $n+1$, since in the appearing new 
terms, the factors $\binom{n}{n+1}$ and $\binom{n}{n+2}$ are zero by 
definition. To unify term~1 and term~3,
we introduce a new index, $k'=k+1$ in term~3 of \eqref{liephir1}:
\begin{equation}
\rho_{ij}^{-2}\sum\limits_{k=0}^{n}\left[-3\binom{n}{k}-2\binom{n}{k+1}\right]L_0^{n-k}\phi_{ij}L_0^{k+1}\Lambda_{ij}= 
\rho_{ij}^{-2}\sum\limits_{k'=1}^{n+1}\left[-3\binom{n}{k'-1}-2\binom{n}{k'}\right]L_0^{n+1-k'}\phi_{ij}L_0^{k'}\Lambda_{ij}. \label{liephir3}
\end{equation}
Note that if we substitute $k'=0$ into the expression after the summation, we
get the same what \eqref{liephir2} is, therefore the latter can be 
inserted into the summation of \eqref{liephir3} while the lower limit
of $k'=1$ is replaced to $k'=0$. Therefore:
\begin{equation}
\% = \rho_{ij}^{-2}\sum\limits_{k'=0}^{n+1}\left[-3\binom{n}{k'}-2\binom{n}{k'+1}-3\binom{n}{k'-1}-2\binom{n}{k'}\right] L_0^{n+1-k'}\phi_{ij}L_0^{k'}\Lambda_{ij} = \%. \label{liephir4}
\end{equation}
Using the relation $\binom{n}{k}+\binom{n}{k+1}=\binom{n+1}{k+1}$,
we get 
\begin{equation}
-3\binom{n}{k}-2\binom{n}{k+1}-3\binom{n}{k-1}-2\binom{n}{k}=
-3\binom{n+1}{k}-2\binom{n+1}{k+1},
\end{equation}
therefore we could simplify \eqref{liephir4}:
\begin{equation}
\% = \rho_{ij}^{-2}\sum\limits_{k=0}^{n+1}\left[-3\binom{n+1}{k}-2\binom{n+1}{k+1}\right]L_0^{n+1-k}\phi_{ij}L_0^{k}\Lambda_{ij}.\label{liephir5}
\end{equation}
Comparing \eqref{liephir5} with \eqref{liephir0}, we conclude that the
relation is proven. For simplicity, we define
\begin{equation}
F_{nk}:=-3\binom{n}{k}-2\binom{n}{k+1}.
\end{equation}
Continuing the derivation of the recurrence formulae, we calculate the
higher order Lie-derivatives of $\Lambda_{ij}$ using the binomial theorem:
\begin{equation}
L_0^n\Lambda_{ij}=\sum\limits_{k=0}^{n}\binom{n}{k}L_0^kA_{ijm}L_0^{n-k}B_{ijm}=
\sum\limits_{k=0}^{n}\binom{n}{k}(L_0^kx_{im}-L_0^kx_{jm})(L_0^{n-k}v_{im}-L_0^{n-k}v_{jm}).
\end{equation}
In the equations of motion the term $\phi_{ij}A_{ijm}$ appears, its
higher order Lie-derivatives can also be calculated like the last relation
for $L_0^n\Lambda_{ij}$:
\begin{equation}
L_0^n(\phi_{ij}A_{ijm})=\sum\limits_{k=0}^{n}\binom{n}{k}L_0^k\phi_{ij}(L_0^{n-k}x_{im}-L_0^{n-k}x_{jm}).
\end{equation}
To summarize our results the complete set of the recurrence relations
for the equations of motion can be found in \eqrefs{nbrec}{nbrec2}.

\section{Derivation of the linearized equations}
\label{appendixrecurrnbodylin}

To obtain the recurrence relations for the linearized equations, we
apply the operator $\Xi\cdot\mathcal{D}$ to \eqrefs{nbrec}{nbrec2}.
We note that the operator $\Xi\cdot\mathcal{D}$ is linear,
\begin{equation}
\Xi\cdot\mathcal{D}(pa+qb)=p(\Xi\cdot\mathcal{D}a)+q(\Xi\cdot\mathcal{D}b),
\end{equation}
where $a$ and $b$ are continuous functions while $p$ and $q$ are constants,
and one can use Leibniz's rule:
\begin{equation}
\Xi\cdot\mathcal{D}(ab)=(\Xi\cdot\mathcal{D}a)b+a(\Xi\cdot\mathcal{D}b).
\end{equation}
Moreover, we should note that the operators $\Xi\cdot\mathcal{D}$
and $L$ cannot be commuted, $\Xi\cdot\mathcal{D}L\ne L\Xi\cdot\mathcal{D}$.

For the first three equations, we get
\begin{eqnarray}
L^{n+1}\xi_{im} & = & \Xi\cdot\mathcal{D}L^{n+1}x_{im} =  \Xi\cdot\mathcal{D}L^{n}v_{im} = L^n\eta_{im}, \\
L^{n}\alpha_{ijm} & = & L^n\xi_{im}-L^n\xi_{jm},  \\
L^{n}\beta_{ijm} & = & L^n\eta_{im}-L^n\eta_{jm}.
\end{eqnarray}
For $\Xi\cdot\mathcal{D}\Lambda_{ij}$ we can use the linear property
and apply Leibniz's rule:
\begin{eqnarray}
\Xi\cdot\mathcal{D}L^n\Lambda_{ij} & = &
\sum\limits_{k=0}^{n}\binom{n}{k}\left[
(\Xi\cdot\mathcal{D}L^kA_{ijm})L^{n-k}B_{ijm}+L^kA_{ijm}(\Xi\cdot\mathcal{D}L^{n
-k}B_{ijm})
\right]=\nonumber \\ 
& = & \sum\limits_{k=0}^{n}\binom{n}{k}\left[
L^k\alpha_{ijm}L^{n-k}B_{ijm}+L^kA_{ijm}L^{n-k}\beta_{ijm})
\right].
\end{eqnarray}
Here we applied the identities 
$\Xi\cdot\mathcal{D}L^{n}A_{ijm}=L^{n}\alpha_{ijm}$
and $\Xi\cdot\mathcal{D}L^{n}B_{ijm}=L^{n}\beta_{ijm}$.
For the calculation of $\Xi\cdot\mathcal{D}L^{n+1}\phi_{ij}$ we follow
the same procedure:
\begin{eqnarray}
\Xi\cdot\mathcal{D}L^{n+1}\phi_{ij} & = &
(\Xi\cdot\mathcal{D})\left[
\rho_{ij}^{-2}\sum_{k=0}^{n}F_{nk}L^{n-k}\phi_{ij}L^k\Lambda_{ij}
\right]=\nonumber \\
& = & (\Xi\cdot\mathcal{D}\rho_{ij}^{-2})(\rho_{ij}^2L^{n+1}\phi_{ij})
+\rho_{ij}^{-2}\sum\limits_{k=0}^{n}F_{nk}
\left[(\Xi\cdot\mathcal{D}L^{n-k}\phi_{ij})L^k\Lambda_{ij}+
L^{n-k}\phi_{ij}(\Xi\cdot\mathcal{D}L^k\Lambda_{ij})\right]. \label{xidphi2}
\end{eqnarray}
The only unknown factor in \eqref{xidphi2} is the 
quantity $\Xi\cdot\mathcal{D}\rho_{ij}^{-2}$. We can calculate it 
easily, because $\rho_{ij}^{-2}$ only depends by definition on the coordinates:
\begin{equation}
\Xi\cdot\mathcal{D}\rho_{ij}^{-2}=\xi_{km}D_{km}(\rho_{ij}^{-2}) =
\xi_{km}(-2)\rho_{ij}^{-3}D_{km}\rho_{ij} =
\left(-2\rho_{ij}^{-3}\right)\xi_{km}D_{km}\rho_{ij} = \%. \label{xidrhopm2}
\end{equation}
The expression $\xi_{km}D_{km}\rho_{ij}$ can be calculated like 
\eqref{nblierho}, where we replace $v_{km}$ by $\xi_{km}$:
\begin{equation}
\xi_{km}D_{km}\rho_{ij}=\rho_{ij}^{-1}(\xi_{im}-\xi_{jm})(x_{im}-x_{jm})=
\rho_{ij}^{-1}\alpha_{ijm}A_{ijm}.
\end{equation}
Thus, we get
\begin{equation}
\%=-2\rho_{ij}^{-4}\alpha_{ijm}A_{ijm}.
\end{equation}
Adding all terms together, we obtain
\begin{equation}
\Xi\cdot\mathcal{D}L^{n+1}\phi_{ij}
=-2\rho_{ij}^{-2}\alpha_{ijm}A_{ijm}L^{n+1}\phi_{ij}
+\rho_{ij}^{-2}\sum\limits_{k=0}^{n}F_{nk}
\left[(\Xi\cdot\mathcal{D}L^{n-k}\phi_{ij})L^k\Lambda_{ij}+
L^{n-k}\phi_{ij}(\Xi\cdot\mathcal{D}L^k\Lambda_{ij})\right].
\end{equation}
The derivation of $L^{n+1}\eta_{im}=\Xi\cdot\mathcal{D}L^{n+1}v_{im}$ is 
the following:
\begin{eqnarray}
L^{n+1}\eta_{im} & = & \Xi\cdot\mathcal{D}L^{n+1}v_{im} =
-G\sum\limits_{j=1, j\ne i}^{K}m_j\left\{\sum\limits_{k=0}^n\binom{n}{k}
\Xi\cdot\mathcal{D}\left(L^k\phi_{ij}L^{n-k}A_{ijm}\right)\right\}= \nonumber \\
& = & -G\sum\limits_{j=1, j\ne i}^{K}m_j 
\left\{\sum\limits_{k=0}^n\binom{n}{k}\left[
(\Xi\cdot\mathcal{D}L^k\phi_{ij})L^{n-k}A_{ijm}+
L^k\phi_{ij}(\Xi\cdot\mathcal{D}L^{n-k}A_{ijm})\right]\right\} = \nonumber \\
& = & -G\sum\limits_{j=1, j\ne i}^{K}m_j 
\left\{\sum\limits_{k=0}^n\binom{n}{k}\left[
(\Xi\cdot\mathcal{D}L^k\phi_{ij})L^{n-k}A_{ijm}+
L^k\phi_{ij}L^{n-k}\alpha_{ijm}\right]\right\}.
\end{eqnarray}
Now we obtained the recurrence relations for all of the linearized
coordinates, velocities and the auxiliary variables 
$\Xi\cdot\mathcal{D}\Lambda_{ij}$, $\Xi\cdot\mathcal{D}\phi_{ij}$. 
The complete set of these equations
are summarized in \eqrefs{nblinrec}{nblinrec2}.

\section{Motion in a reference frame fixed to one of the bodies}
\label{appendixfixednb}

Throughout the derivation of the recurrence relations, we can use the fact
that the partial differential operators $\frac{\partial}{\partial r_{im}}$ 
and $\frac{\partial}{\partial w_{im}}$ are equivalent to 
$D_{im}=\frac{\partial}{\partial x_{im}}$ and 
$\Delta_{im}=\frac{\partial}{\partial v_{im}}$, because the variables
differ only in a constant ($x_{0m}$ and $v_{0m}$, respectively).
We have to define the new variable $\Lambda_i=r_{im}w_{im}$.
The derivation of the recurrence relations can be done following
the steps of \apref{appendixrecurrnbody} and 
\apref{appendixrecurrnbodylin}: the quantities $\rho_{i}$,
$\phi_{i}$ and $\Lambda_{i}$ have the same properties for the Lie-derivation
as $\rho_{ij}$, $\phi_{ij}$ and $\Lambda_{ij}$, respectively, therefore
all of the induction steps can be done in the appropriate way.

Thus, the recurrence relations
for the $N$-body problem around a fixed centre can be written as
\begin{eqnarray}
L^{n+1}r_{im} & = & L^nw_{im}, \label{fixnbrec}\\
L^{n}A_{ijm} & = & L^nr_{im}-L^nr_{jm},  \\
L^{n}B_{ijm} & = & L^nw_{im}-L^nw_{jm},  \\
L^{n}\Lambda_{i} & = & \sum\limits_{k=0}^{n}\binom{n}{k}L^kr_{im}L^{n-k}w_{im},  \\
L^{n}\Lambda_{ij} & = & \sum\limits_{k=0}^{n}\binom{n}{k}L^kA_{ijm}L^{n-k}B_{ijm},  \\
L^{n+1}w_{im} & = & -G(\mathcal{M}+m_i)\sum\limits_{k=0}^n\binom{n}{k}L^k\phi_iL^{n-k}r_{im} - G\sum\limits_{j=1, j\ne i}^{K}m_j\sum\limits_{k=0}^n\binom{n}{k}\left[L^k\phi_{ij}L^{n-k}A_{ijm}+L^k\phi_jL^{n-k}r_{jm}\right], \\
L^{n+1}\phi_{i} & = & \rho_{i}^{-2}\sum\limits_{k=0}^{n}F_{nk}L^{n-k}\phi_{i}L^k\Lambda_{i},  \\
L^{n+1}\phi_{ij} & = & \rho_{ij}^{-2}\sum\limits_{k=0}^{n}F_{nk}L^{n-k}\phi_{ij}L^k\Lambda_{ij}. \label{fixnbrec2}
\end{eqnarray}
Let us denote the linearized of $r_{im}$ and $w_{im}$ by $\xi_{im}$ and 
$\eta_{im}$, respectively. Since
$\alpha_{ijm}=\xi_{im}-\xi_{jm}$ and $\beta_{ijm}=\eta_{im}-\eta_{jm}$,
for the linearized equations the calculations yield
\begin{eqnarray}
L^{n+1}\xi_{im} & = & L^n\eta_{im}, \label{fixnblinrec}\\
L^{n}\alpha_{ijm} & = & L^n\xi_{im}-L^n\xi_{jm},  \\
L^{n}\beta_{ijm} & = & L^n\eta_{im}-L^n\eta_{jm}, \\
\Xi\cdot\mathcal{D}L^{n}\Lambda_{i} & = & \sum\limits_{k=0}^{n}\binom{n}{k}\left(L^k\xi_{im}L^{n-k}w_{im}+L^kr_{im}L^{n-k}\eta_{im}\right), \\
\Xi\cdot\mathcal{D}L^{n}\Lambda_{ij} & = & \sum\limits_{k=0}^{n}\binom{n}{k}\left(L^k\alpha_{ijm}L^{n-k}B_{ijm}+L^kA_{ijm}L^{n-k}\beta_{ijm}\right),  \\
L^{n+1}\eta_{im} & = & -G(\mathcal{M}+m_i)\sum\limits_{k=0}^n\binom{n}{k}\left[(\Xi\cdot\mathcal{D}L^k\phi_i)L^{n-k}r_{im}+L^k\phi_iL^{n-k}\xi_{im}\right] - \nonumber \\
                & & - G\sum\limits_{j=1, j\ne i}^{K}m_j\sum\limits_{k=0}^n\binom{n}{k}\left[(\Xi\cdot\mathcal{D}L^k\phi_{ij})L^{n-k}A_{ijm}+L^k\phi_{ij}L^{n-k}\alpha_{ijm}+(\Xi\cdot\mathcal{D}L^k\phi_j)L^{n-k}r_{jm}+L^k\phi_jL^{n-k}\xi_{jm}\right],  \\
\Xi\cdot\mathcal{D}L^{n+1}\phi_{i} & = & -2\rho_{i}^{-2}\xi_{im}r_{im}L^{n+1}\phi_{i}+ \rho_{i}^{-2}\sum\limits_{k=0}^{n}F_{nk}\left[(\Xi\cdot\mathcal{D}L^{n-k}\phi_{i})L^k\Lambda_{i}+L^{n-k}\phi_{i}(\Xi\cdot\mathcal{D}L^k\Lambda_{i})\right],  \\
\Xi\cdot\mathcal{D}L^{n+1}\phi_{ij} & = & -2\rho_{ij}^{-2}\alpha_{ijm}A_{ijm}L^{n+1}\phi_{ij} + \rho_{ij}^{-2}\sum\limits_{k=0}^{n}F_{nk}\left[(\Xi\cdot\mathcal{D}L^{n-k}\phi_{ij})L^k\Lambda_{ij}+L^{n-k}\phi_{ij}(\Xi\cdot\mathcal{D}L^k\Lambda_{ij})\right]. \label{fixnblinrec2}
\end{eqnarray}

\section{Speed-up considerations}
\label{appendixspeedup}

Introducing new variables, the required number of arithmetical operations
can be decreased in \eqrefs{fixnbrec}{fixnbrec2} and \eqrefs{fixnblinrec}{fixnblinrec2}. Namely,
the calculation of $L^{n+1}w_{im}$ and $L^{n+1}\eta_{im}$ can be written as
\begin{eqnarray}
L^{n+1}w_{im} & = & -G(\mathcal{M}+m_i)S^{[n]}_{im}-G\sum\limits_{j=1, j\ne i}^{K}m_j\left(S^{[n]}_{ijm}+S^{[n]}_{jm}\right), \\
L^{n+1}\eta_{im} & = & -G(\mathcal{M}+m_i)\Sigma^{[n]}_{im}-G\sum\limits_{j=1, j\ne i}^{K}m_j\left(\Sigma^{[n]}_{ijm}+\Sigma^{[n]}_{jm}\right), 
\end{eqnarray}
where the new variables are
\begin{eqnarray}
S^{[n]}_{im} & = & \sum\limits_{k=0}^n\binom{n}{k}L^k\phi_i L^{n-k}r_{im}, \\
S^{[n]}_{ijm} & = & \sum\limits_{k=0}^n\binom{n}{k}L^k\phi_{ij} L^{n-k}A_{ijm}, \\
\Sigma^{[n]}_{im} & = & \sum\limits_{k=0}^n\binom{n}{k}\left[(\Xi\cdot\mathcal{D}L^k\phi_i)L^{n-k}r_{im}+L^k\phi_iL^{n-k}\xi_{im}\right], \\
\Sigma^{[n]}_{ijm} & = & \sum\limits_{k=0}^n\binom{n}{k}\left[(\Xi\cdot\mathcal{D}L^k\phi_{ij})L^{n-k}A_{ijm}+L^k\phi_{ij}L^{n-k}\alpha_{ijm}\right]. 
\end{eqnarray}
The implementation of the above relations can increase the speed of the
calculations by 20\%-30\%, depending on the number of the bodies.

\bsp

\label{lastpage}

\end{document}